\documentclass[floatfix,notitlepage,twocolumn,longbibliography]{revtex4-1}%
\usepackage{graphicx}%
\usepackage{amsmath}%
\usepackage{amsfonts}%
\usepackage{amssymb}
\usepackage{bm}
\usepackage{comment}
\usepackage{color}
\usepackage[breaklinks]{hyperref}
\usepackage[normalem]{ulem}

\def\d{{\partial}}
\def\s{{\sigma}}

\def\A{{ {\bm A} }}
\def\k{{ {\bm k} }}
\def\p{{ {\bm p} }}
\def\q{{ {\bm q} }}

\def\B{{ {\bm B} }}

\def\0{{ {\bm 0} }}
\def\w{{\omega}}
\def\a{{\alpha}}
\def\b{{\beta}}							
\def\g{{\gamma}}

\def\bpsi{\bar{\psi}}
\def\ve{{\varepsilon}}
\def\r{{ {\bm r} }}
\def\rK{{ {\rm K} }}
\def\rR{{ {\rm R} }}
\def\rA{{ {\rm A} }}
\def\expo{{ {\rm e} }}
\def\cV{{ {\cal V} }}
\def\GRmGA{{ (\hat{G}^\rR - \hat{G}^\rA) }}
\def\ral{{  \ell \rangle}}
\def\lal{{ \langle \ell  }}
\def\ran{{  n \rangle }}
\def\lan{{ \langle n  }}
\def\ram{{ m \rangle }}
\def\lam{{ \langle m  }}
\def\daran{{  \partial_\a n \rangle }}
\def\dalan{{ \langle \partial_\a n  }}
\def\daram{{  \partial_\a m \rangle }}

\def\dblan{{ \langle \partial_\b n  }}
\def\dbram{{  \partial_\b m \rangle }}

\usepackage{xcolor}

\allowdisplaybreaks[4]

\begin{document} 
\title{Interband contributions to nonlinear transport in semiconductor nanostructures}
\author{ 
Kazuki Nakazawa$^1$,  
Henry F. Legg$^2$, 
Jelena Klinovaja$^2$, and 
Daniel Loss$^{1,2}$ } 
\address{
$^1$RIKEN Center for Emergent Matter Science (CEMS), 2-1 Hirosawa, Wako, Saitama 351-0198, Japan
\\
$^2$Department of Physics, University of Basel, Klingelbergstrasse 82, 4056 Basel, Switzerland}

\date{\today}

\begin{abstract}
Spin-orbit interaction (SOI) is a crucial ingredient for many potential applications of quantum devices, such as the use of semiconductor nanostructures for quantum computing. It is known that nonlinear conductivities are sensitive to the strength and type of SOI, however, many calculations of nonlinear transport coefficients are based on the semiclassical Boltzmann theory and make simplifying assumptions about scattering effects due to disorder. In this paper, we develop and employ a microscopic theory based on the Keldysh formalism that goes beyond simple semiclassical approximations. This approach, for instance, naturally takes into account the effects of interband transitions, Berry curvature, and allows for a more precise treatment of impurity scattering. As a test of this formalism, we consider the nonlinear transport properties in an effective two-band model of one-dimensional nanowires (1DNWs) and two-dimensional hole gases (2DHGs) in the presence of a magnetic field causing Zeeman splitting of the spin states. We find that the small energy scales in nanostructures can result in interband contributions that are relevant, especially in the dirty limit, and therefore could modify qualitative features found using a purely semiclassical approach. Nonetheless, we find that different types of SOI (linear or cubic) still result in remarkably pronounced different in-plane field angle dependences, which survive even when interband effects are relevant. Our results provide a detailed understanding of when interband effects become important for nonlinear transport and can serve as the basis for a microscopic description to predict other nonlinear transport effects in materials and devices.
\end{abstract}

\maketitle

\section{Introduction}
\label{sec:intro}

The utility of transport phenomena comes largely from the possibility to detect non-trivial electronic and magnetic structures via, e.g., the anomalous Hall effect~\cite{Nagaosa2010a}, Kondo effect~\cite{Kondo1964,AFL1983}, (giant, tunneling) magnetoresistance~\cite{Baibich1988,Binasch1989}, and topological Hall effect~\cite{LG1992,Ye1999,TK2002,Bruno2004a,Neubauer2009,NBK2018,NK2019}. However, the vast majority of well established transport probes are linear response phenomena, in the sense that they are proportional to the first order of the driving field, e.g., an external electric field. 

Nonlinear transport phenomena are responses proportional to higher orders of the external driving field and have been of increasing interest in recent years. In particular, the breaking of the spatial inversion symmetry of a system often plays an essential role in the manifestation of the second-order response to an external field~\cite{TSM2018,Ma2019,CSS2022,NKM2022,Yokouchi2017,IN2020,OMKM2021,SF2015,DLX2021,DWSLX,Ideue2017,He2018,Legg2022,Wang2022,LLK2022,Wang2023,YNY2023,Pan2019}, i.e., in the presence of conical magnetism~\cite{Yokouchi2017,IN2020,OMKM2021}, Berry curvature dipoles ~\cite{SF2015,DLX2021,DWSLX}, and polar/chiral effects~\cite{Ideue2017,He2018,Legg2022,Wang2022,Guo2022,LLK2022,Wang2023,YNY2023}. Due to their potential to provide novel probes of systems, the foundations of nonlinear response phenomena have recently been studied intensively. This includes microscopic formulations~\cite{PMOM,JL,MP,MN,YNY2023} where the correspondence with quantum geometry~\cite{GYN2014,MN}, generalized Berry curvature~\cite{MN}, and orbital magnetic moments~\cite{YNY2023} has been investigated. 
    
The links between nonlinear transport phenomena and inversion symmetry makes them ideal probes of spin-orbit interaction (SOI) and effects that result from SOI~\cite{Ideue2017,Legg2022,DLBLK}. Many phenomena originating from SOI are known  (e.g., spin Hall effect~\cite{DP1971PLA,DP1971JETP,MNZ2003,Sinova2004,CL2005,CL2009,DML2009}, spin-orbit torque~\cite{Miron2011,Liu2012,Manchon2019,Shao2021}, and magnetic skyrmion-related phenomena~\cite{Muhlbauer2009,Heinze2011,Kiselev_2011,NT2013,YSKL2016,FPHLE2016,PHKL2017}) and these have been vigorously studied, e.g., for their applications to spintronic devices~\cite{ZFS2004,Hirohata2020}. In addition, the engineering of SOI in semiconductor nanostructures is of paramount importance in quantum information processing~\cite{LD1998,HKPTV2007,KL2013}. In certain nanostructures, such as nanowires and two-dimensional heterostructures, SOI can be controlled by electric fields due to gating~\cite{KTL2011,KRL2018,ABBKL2022,ABKL2022,Gao2020,VBPN2018,MVN2021,Bellentani2021}.  Notably, two-dimensional hole gases (2DHG) in planar germanium heterostructures have drawn significant attention for their potential in quantum information processing~\cite{Scappucci2021,Hendrickx2020a,Hendrickx2020b,Hendrickx2021}. 
   
Recently, it has been proposed that nonlinear transport can be used to determine SOI in nanostructures~\cite{DLBLK,Huang2023}. In particular, it was shown that transport coefficients second-order in electric field differ significantly depending on whether the SOI is linear or cubic in nature~\cite{DLBLK}. This difference in nonlinear response enables a characterization of the type and strengths of SOI early in the device fabrication process. However, nonlinear responses are often calculated using semiclassical Boltzmann transport theory, within which it is generally difficult to fully incorporate effects such as interband transitions, which could be important on the small energy scales found in nanostructures. In addition, this semiclassical approach often employs simple phenomenological assumptions about the details of the system, such as the momentum relaxation of charge carriers. A microscopic description of nonlinear transport effects, going beyond a simple phenomenological description, would therefore enable a fuller characterization of, e.g., the SOI in nanostructures used to realise hole qubits. 

In this paper, we investigate the effects of interband transitions in nonlinear transport and employ a detailed description of momentum relaxation due to impurity scattering. We start from the ground up by establishing a microscopic description of nonlinear effects based on the Keldysh formalism. We then apply our results to semiconductor nanostructures such as one-dimensional nanowires (1DNWs) and two-dimensional  hole gases (2DHGs). First, we demonstrate that the corresponding intraband terms in the microscopic expression are consistent with Boltzmann theory. 
Based on the band representation of the energy dispersion, 
we also discuss the various contributions to nonlinear transport behavior. Following this, we apply our microscopic formalism to low-dimensional nanostructures as a testbed for the new effects incorporated in this description. In the case of 1DNWs and 2DHGs, we show that interband contributions increase near the band crossing and discuss their relevance for characterising SOI in these nanostructures. Furthermore, we also show that, when the self-energy is treated within the Born approximation or self-consistent Born approximation, the van Hove singularities in 1DNWs and the off-diagonal (non-spin conserving) component of the self-energy can, in certain circumstances, significantly affect the behavior of second-order nonlinear conductivity. Overall, our microscopic analysis provides a detailed understanding of when interband and impurity scattering effects can have an influence on nonlinear transport properties beyond simple semiclassical approaches. Furthermore, our results provide the basis for a full microscopic description that can be used to predict a variety of nonlinear transport effects in materials and devices. 
    
This paper is organized as follows. In Sec.~\ref{sec:formalism}, we introduce the general formalism to obtain the second-order nonlinear conductivity, as well as describe the interband and intraband contributions. In Sec.~\ref{sec:1D} and \ref{sec:2D}, we discuss the second-order responses in the effective model of 1DNW and 2DHG under the magnetic field. In Sec.~\ref{sec:discussion}, we discuss the results obtained in the previous sections and provide messages for future measurements. Sec.~\ref{sec:summary} provides a summary of our results.

\section{Formalism}
\label{sec:formalism}

\begin{figure*}[t]
  \includegraphics[width=175mm]{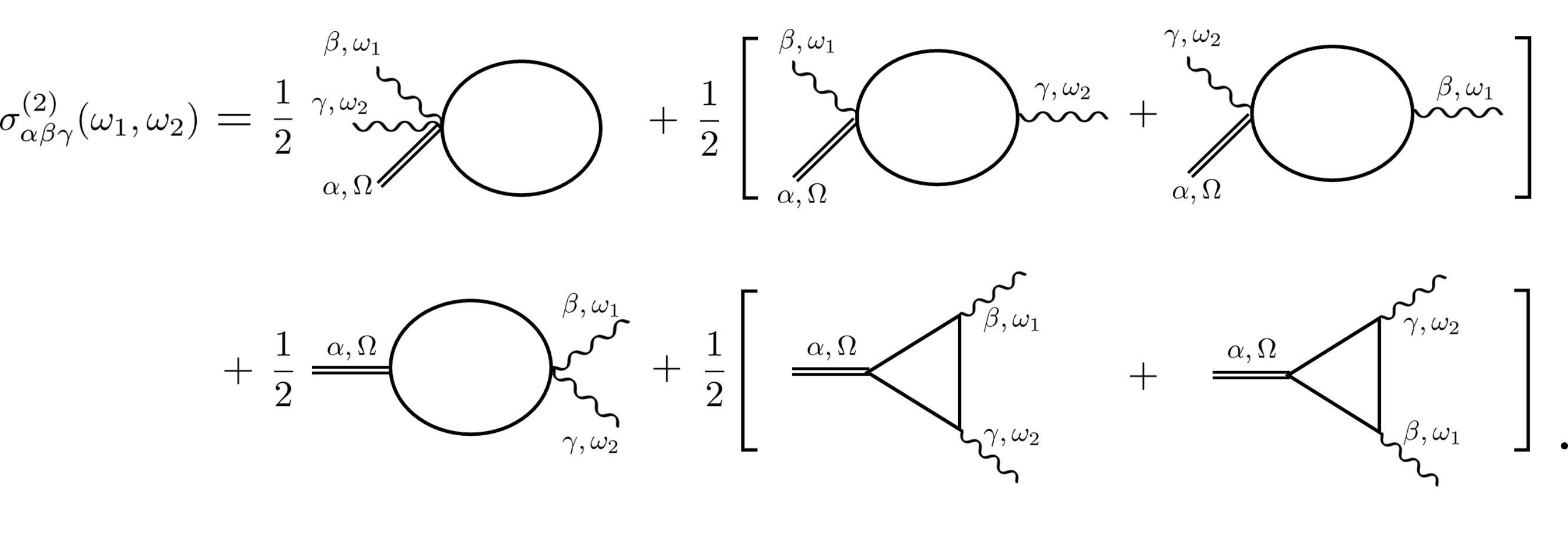}
 \caption{ 
Diagrammatic representation  of the second-order optical conductivity Eq.~(\ref{eq:ja}). Wavy lines and double lines in the Feynman diagrams represent the electromagnetic coupling and response current, respectively. The analytical expressions of the Green functions (solid lines) are not specified in this notation. The last three diagrams are responsible for the second-order DC conductivity Eq.~\eqref{eq:DC}. 
}
 \label{fig:K2}
\end{figure*}

\subsection{General formula for nonlinear conductivities}
\label{subsec:General}

To derive the DC second-order nonlinear conductivity, we here employ the Keldysh Green functions based on the path integral formalism~\cite{Kamenev}. Throughout we will set $\hbar = 1$, unless otherwise stated. The electric current $j_i$ can be calculated using the current operator $\hat{j}_i$:
\begin{align}
\langle \hat{j}_i \rangle = -\frac{i}{2} {\rm tr} [ \hat{j}_i \hat{G}^{\rm K} ],
\label{eq:j}
\end{align}
where $\hat{G}^\rK$ is a Keldysh Green function, and $\rm tr$ is the trace taken over all degrees of freedom (e.g., momentum, frequency, spin, and orbital).

To discuss the response to the electric field, let us assume a continuum model described by the Hamiltonian $\hat{\cal{H}} = \hat{H} (\p - e\A) + \hat{H}_{\rm i}$ with ${\p=-i \nabla}$ being the momentum operator, $\A \equiv {\bm A} (t)$ the time-dependent $U$(1) vector potential which is assumed to be uniform in real space, and  $\hat{H}_{\rm i}$ the impurity scattering term. The hat on the Hamiltonian indicates that it is a matrix with respect to discrete degrees of freedom (e.g., for spin and orbital band index). Let us consider a Fourier component $\hat{H} (\k - e\A)$ where $\k$ is the wave vector, and expand it with respect to $\A$, 
\begin{align}
\hat{H} (\k - e{\bm A}) \simeq \hat{H}_\k &- eA_\a \partial_\a \hat{H}_\k + \frac{e^2}{2} A_\a A_\b \partial_\a \partial_\b \hat{H}_\k \nonumber \\
                                    &- \frac{e^3}{6} A_\a A_\b A_\g \partial_\a \partial_\b \partial_\g \hat{H}_\k, 
\end{align}
leading to a current operator of the form, 
\begin{align}
&\hat{j}_\a (\k) = \frac{ \delta \hat{H} (\k - e{\bm A})}{\delta A_\a} \nonumber \\ 
            &\simeq  - e \partial_\a \hat{H}_\k + e^2 A_\b \partial_\a \partial_\b \hat{H}_\k 
            - \frac{e^3}{2} A_\b A_\g \partial_\a \partial_\b \partial_\g \hat{H}_\k, 
\end{align}
where $e$ is a charge and $\partial_i = \partial/\partial k_i$. From here on we define $\hat{\cV}_{\a_1 \a_2 \cdots \a_n} \equiv \partial_{\a_1} \partial_{\a_2} \cdots \partial_{\a_n} \hat{H}_\k$. Here, $\hat{H}_\k \equiv \hat{H} (\k - e{\bm A})|_{{\bm A}=0}$ is the momentum space representation of the Hamiltonian without the coupling to the external field, and we define the retarded and advanced Green functions $\hat{G}_{\k\k'}^{\rR} (\ve) = (\ve - \hat{\cal H}|_{\A = 0} + i0^+)_{\k\k'}^{-1} = (\hat{G}_{\k\k'}^\rA)^\dagger$, which takes the full account of the system including the impurity scattering which we discuss in Sec.~\ref{sec:self}. Thanks to this expansion we can obtain the response to the general order of the vector potential ${\bm A}$. In particular, we are interested in the second-order response of the electric field, given in frequency space $\w$ by ${\bm E} (\w)= i\w {\bm A} (\w)$. After first calculating the nonlinear conductivity within the Kelydsh framework, we will then check the consistency with  previous studies~\cite{PMOM,JL,MP} (see Appendix~\ref{sec:general_detail}). The diagrams corresponding to the different nonlinear contributions are shown in Fig.~\ref{fig:K2}. Substituting these operators into Eq.~(\ref{eq:j}) and retain the second order of the vector potentials and frequencies $\w_1$ and $\w_2$, while taking the $\w_1, \w_2 \to 0$ limit to consider the uniform and static (DC) electric field, we obtain 
\begin{align}
\langle \hat{j}_\a^{(2)} \rangle   &= \sigma_{\a \b \g} E_\b E_\g  \nonumber \\  &= ( \sigma_{\a\b\g}^{\rm RRA} + \sigma_{\a\b\g}^{\rm RA} + \sigma_{\a\b\g}^{\rm RRR} + \sigma_{\a\b\g}^{\rm RR} ) E_\b E_\g , \label{eq:DC}  \\
\sigma_{\a\b\g}^{\rm RRA} &=-e^3{\rm Im} \ {\rm tr}  \Biggl[ \left(-\frac{\partial f}{\partial \ve}\right) \hat{\cV}_\a \frac{\partial \hat{G}^\rR }{\partial \ve} \hat{\cV}_\b \hat{G}^\rR \hat{\cV}_\g \hat{G}^\rA \Biggr], \nonumber 
\\
                                         & \quad + (\b \leftrightarrow \g)  \label{eq:RRA}  
                                         \\  
\sigma_{\a\b\g}^{\rm RA}   &=-e^3 {\rm Im} \ {\rm tr}  \Biggl[ \left(-\frac{\partial f}{\partial \ve}\right) \hat{\cV}_\a \frac{\partial \hat{G}^\rR }{\partial \ve} \hat{\cV}_{\b\g} \hat{G}^\rA \Biggr], \label{eq:RA}  
\\
\sigma_{\a\b\g}^{\rm RRR} &= -e^3 {\rm Im} \ {\rm tr} \Biggl[ f(\ve) \hat{\cV}_\a \frac{\partial}{\partial \ve} \left( \frac{\partial \hat{G}^\rR }{\partial \ve} \hat{\cV}_\b \hat{G}^\rR \right) \hat{\cV}_\g \hat{G}^\rR \Biggr], \nonumber \\
                                         &\quad + (\b \leftrightarrow \g) \label{eq:RRR} 
                                         \\
\sigma_{\a\b\g}^{\rm RR}   &= -e^3 {\rm Im} \ {\rm tr} \Biggl[ f(\ve) \frac{\partial^2 \hat{G}^\rR }{\partial \ve^2} \hat{\cV}_{\b\g} \hat{G}^\rR \Biggr] .  \label{eq:RR} 
\end{align}
Equations~\eqref{eq:DC}-\eqref{eq:RR} are consistent with Ref.~\citenum{MP}. In particular, $\sigma_{\a\b\g}^{\rm RRA(RRR)}$ correspond to the last two Feynman diagrams and $\sigma_{\a\b\g}^{\rm RA(RR)}$ corresponds to the fourth diagram in Fig.~\ref{fig:K2}. To be physical in the DC limit, the terms proportional to $\w_i^n/\w_1 \w_2$ in $\sigma_{\a\b\g}$ should be absent, since they would lead to divergences in the limit $\lim_{\w_1,\w_2 \to 0} \sigma_{\a \b \g} (\w_1,\w_2)$. We indeed confirm the absence of the divergent terms in Appendix~\ref{sec:DClimit}.

Hereafter, we assume  zero temperature, put the Fermi function $f(\ve) = \theta (\mu - \ve)$, with $\mu$ the chemical potential, and only consider the $\sigma_{\a\b\g}^{\rm RRA}$ and $\sigma_{\a\b\g}^{\rm RA}$ terms since we only consider a metallic region which has a Fermi surface. Note that the terms $\sigma_{\a\b\g}^{\rm RR(R)}$ will be subdominant in the the hole relaxation time (caused by static random disorder) $\tau$ (see below) and so can be disregarded when a magnetic field breaks  time reversal symmetry~\cite{MN}, which is the regime of main interest in this work.

\subsection{Self-energy}
\label{sec:self}

The effect of elastic scattering due to the impurities is incorporated by considering the self-energy $\Sigma$ of the impurity-averaged Green function $\hat{G}_\k^\rR (\ve) = ( \ve - \hat{H}_\k - \hat{\Sigma}^\rR (\ve) )^{-1}$. The simplest approximation is to put the constant damping rate $\gamma$ as the self-energy $\Sigma^\rR = -i\g \equiv -\frac{i\hbar}{2\tau}$, where $\tau$ is the momentum relaxation time. This is an especially good approximation in the non-interacting two-dimensional electron (hole) gases as the density of states are approximately constant in energy in 2D, see below. 

To go further we can also include a more microscopic description of the electron damping. For instance, the potential from point like static impurities can be described as 
\begin{align}
\hat{H}_{\rm i} = u 
\sum_{j=1}^{N_{\rm i}}  \delta ({\bm r} - {\bm X}_j) ,
\end{align} 
where $u$, $N_{\rm i}$, and ${\bm X}_j$ are the strength of the impurity scattering, the number of impurities, and the random position of the $j$th impurity, respectively.  
The self-energy with respect to the impurity scattering can be treated within the Born approximation, 
\begin{align}
\hat{\Sigma}^\rR (\ve) = 
\frac{n_{\rm i} u^2}{V} \sum_{\bm k} \hat{g}_\k^\rR (\ve), 
\end{align}
where $\hat{g}_\k^{\rm R} (\ve) = ( \ve - \hat{H}_{\bm k} + i0^+ ) ^{-1}$ is the bare retarded Green function and $n_{\rm i} = N_{\rm i} / V$ is the impurity concentration with the system volume $V$. A more accurate treatment can be achieved by considering the self-consistent treatment of the self-energy within the Born approximation, referred to as self-consistent Born approximation (SCBA):
\begin{align}
\hat{\Sigma}_{\rm SCBA}^{\rR} (\ve) = 
\frac{n_{\rm i} u^2}{V} \sum_{\bm k} \hat{G}_{\bm k}^\rR (\ve), 
\end{align}
where the self-energy is self-consistently defined, such that one has to perform the summation iteratively until convergence.

\subsection{Intraband and Interband terms}
\label{sec:bandrep}

The simplest way to investigate the nonlinear transport properties is to apply Boltzmann transport theory. For this we use the eigenvalue equation for $\hat{H}_\k$, $\hat{H}_\k | n \k \rangle = \ve_{n \k} | n \k \rangle$, with eigenenergy  $\ve_{n\k}$ and eigenket $| n \k \rangle$, where $n$ denotes the band index. Expanding up to the second order of electric field within the relaxation time approximation and at zero temperature, we get~\cite{Legg2022,Wang2022,DLBLK} 
\begin{align}
\sigma_{\a\b\g}^{\rm B} = - e^3 \tau^2  
\sum_\k \sum_{n} 
v_\a^{(n)} v_{\b\g}^{(n)} \delta (\mu - \ve_{n\k}) ,
\label{eq:Boltz}
\end{align} 
where, again, $\tau$  represents the momentum relaxation time, and $v_\a^{(n)} = \partial_\a \ve_{n\k}$ and $v_{\a\b}^{(n)} = \partial_\a \partial_\b  \ve_{n\k}$ are the group velocity and band curvature, respectively. Note that $\s_{\a\b\g}^{\rm B}$ only contains the intraband contributions. 

We now derive the band representation from the microscopic formula Eq.~(\ref{eq:DC}) for comparison to the Boltzmann formalism and the physical descriptions. For this, let us, for simplicity, consider a constant self-energy. Also, as the eigenvalues, eigenstates, and the Green functions are characterized in each wave number $\k$ and the Green functions are evaluated at $\ve = \mu$, we hereafter use the notations $\ve_n \equiv \ve_{n\k}$, $|n \rangle \equiv | n\k \rangle$, and  $G_n^{\rR (\rA)} \equiv G_{\k n}^{\rR (\rA)} (\mu) $ in the diagonal basis. Thus, we can write 
\begin{align}
G_n^{\rR (\rA)}\equiv \langle n | \hat{G}_\k^{\rR (\rA)} | n \rangle . 
\label{eq:uni}
\end{align} 
Introducing the notation $\ve_{nm} \equiv \ve_n - \ve_m$, the current operator and its derivative are modified such that 
\begin{align}
\lan | \hat{\cV}_\a | \ram &= \partial_\a \ve_n \delta_{nm} + i \ve_{mn} A_\a^{nm} ,
\label{eq:UAU}
\\
\lan | \hat{\cV}_{\a\b} | \ram &= \frac{1}{2} \partial_\a \partial_\b \ve_n \delta_{nm} + \frac{i}{2} \ve_{nm} \partial_\b A_\a^{mn}  \nonumber \\
&+ i \partial_\b \ve_{nm} A_\a^{mn}  \nonumber \\
&- \sum_\ell \ve_{n\ell} A_\a^{m\ell} A_\b^{\ell n}  + (\a \leftrightarrow \b) , 
\label{eq:UABU}
\end{align}  
where $A_\a^{nm} = -i \lan | \daram$. The first terms in Eqs.~(\ref{eq:UAU}) and (\ref{eq:UABU}) are the intraband components, and all other terms express the interband processes. Hence we can decompose $\sigma_{\a\b\g}^{\rm RRA}$ and $\sigma_{\a\b\g}^{\rm RA}$ into the interband and intraband contributions,   
\begin{align}
\sigma_{\a\b\g}^{\rm RRA} &\equiv
 \sigma_{\a\b\g}^{\rm RRA, intra} + \sigma_{\a\b\g}^{\rm RRA, inter},   
\label{eq:decRRA}
\\
\sigma_{\a\b\g}^{\rm RA} &\equiv
 \sigma_{\a\b\g}^{\rm RA, intra} + \sigma_{\a\b\g}^{\rm RA, inter} , 
 \label{eq:decRA}
\end{align}
and define $\s_{\a\b\g}^{\rm intra} = \sigma_{\a\b\g}^{\rm RRA, intra} + \sigma_{\a\b\g}^{\rm RA, intra}$ and $ \sigma_{\a\b\g}^{\rm inter} =  \sigma_{\a\b\g}^{\rm RRA, inter} + \sigma_{\a\b\g}^{\rm RA, inter}$, where the explicit definitions of these terms are given  in Appendix~\ref{sec:bandrep}. 

Let us first comment on $\sigma_{\a\b\g}^{\rm intra}$. This term can be expressed as 
\begin{align}
\sigma_{\a\b\g}^{\rm intra} =  \frac{e^3}{\pi} \tau^2 \sum_\k \sum_n v_\a^{(n)} v_{\b\g}^{(n)} {\rm Im} G_{n}^\rR . 
\label{eq:intrarel}
\end{align}
We see that this intraband term is proportional to $\tau^2$ and ${\rm Im} G_n^\rR \to -\pi \delta(\mu - \ve_n)$ in the $\g \to 0$ limit, this intraband contribution is analogous to Eq.~(\ref{eq:Boltz}) obtained from the Boltzmann theory. 

Next, we consider the interband term $\sigma_{\a\b\g}^{\rm inter}$. The band expression of $\sigma_{\a\b\g}^{\rm inter}$ is quite complicated, as it contains multifarious interband processes involving a large number of terms (see Appendix~\ref{sec:bandrep}). To get some insights on the disorder dependence, let us consider the dirty limit, such that $\ve_{nm}\tau \ll 1$ for all wave numbers. In this limit we can write $\sigma_{\a\b\g}^{\rm inter} \equiv \sigma_{\a\b\g}^{\rm BCD} + \sigma_{\a\b\g}^{\rm QM}$ to  leading order in $\ve_{nm} \tau$, where 
\begin{align}
\sigma_{\a\b\g}^{\rm BCD} 
&\simeq \frac{e^3}{\pi} \tau \sum_\k \sum_{n} v_\b^{(n)} \ve_{\a \g \delta} \Omega_\delta^{(n)} {\rm Im} G_{n}^\rR  
 + \b \leftrightarrow \g,
\label{eq:BCD}
\\
\sigma_{\a\b\g}^{\rm QM} 
&\simeq \frac{e^3}{\pi} \tau^2 \sum_\k \sum_{n\ell} \ve_{n\ell} g_{\a ; \b\g}^{n\ell} {\rm Im} G_{n}^\rR .
\label{eq:domQM}
\end{align} 
Here, $\Omega_\delta^{(n)}$ is the Berry curvature and $g_{\a;\b\g}^{n\ell}$ is a symmetric tensor with respect to the band indices $n$ and $m$ with $\ve_{\a\g\delta}$ the Levi-Civita symbol (see Appendix~\ref{sec:bandrep} for details). The first term Eq.~\eqref{eq:BCD} is the Berry curvature dipole term, which is proportional to $\tau$ and responsible for the nonlinear Hall effect~\cite{SF2015,DLX2021,DWSLX}. We can also obtain this expression without requiring the criterion $\ve_{nm}\tau \ll 1$ (see Appendix~\ref{sec:bandrep}). The second term, Eq.~\eqref{eq:domQM}, contains the rest of the interband term, which is related to the quantum metric~\cite{GYN2014,MN} specified by $g_{\a ;\b\g}^{n\ell}$. This term has a $\tau^2$-dependence in the leading order if we assume $\ve_{nm}\tau \ll 1$. This limit provides insight into the origin of the various interband contributions, in later sections, we will show that this approximation works particularly well for dirty systems with small scattering times $\tau$, but higher order corrections can be important in cleaner systems.

\begin{figure*}
\includegraphics[width=175mm]{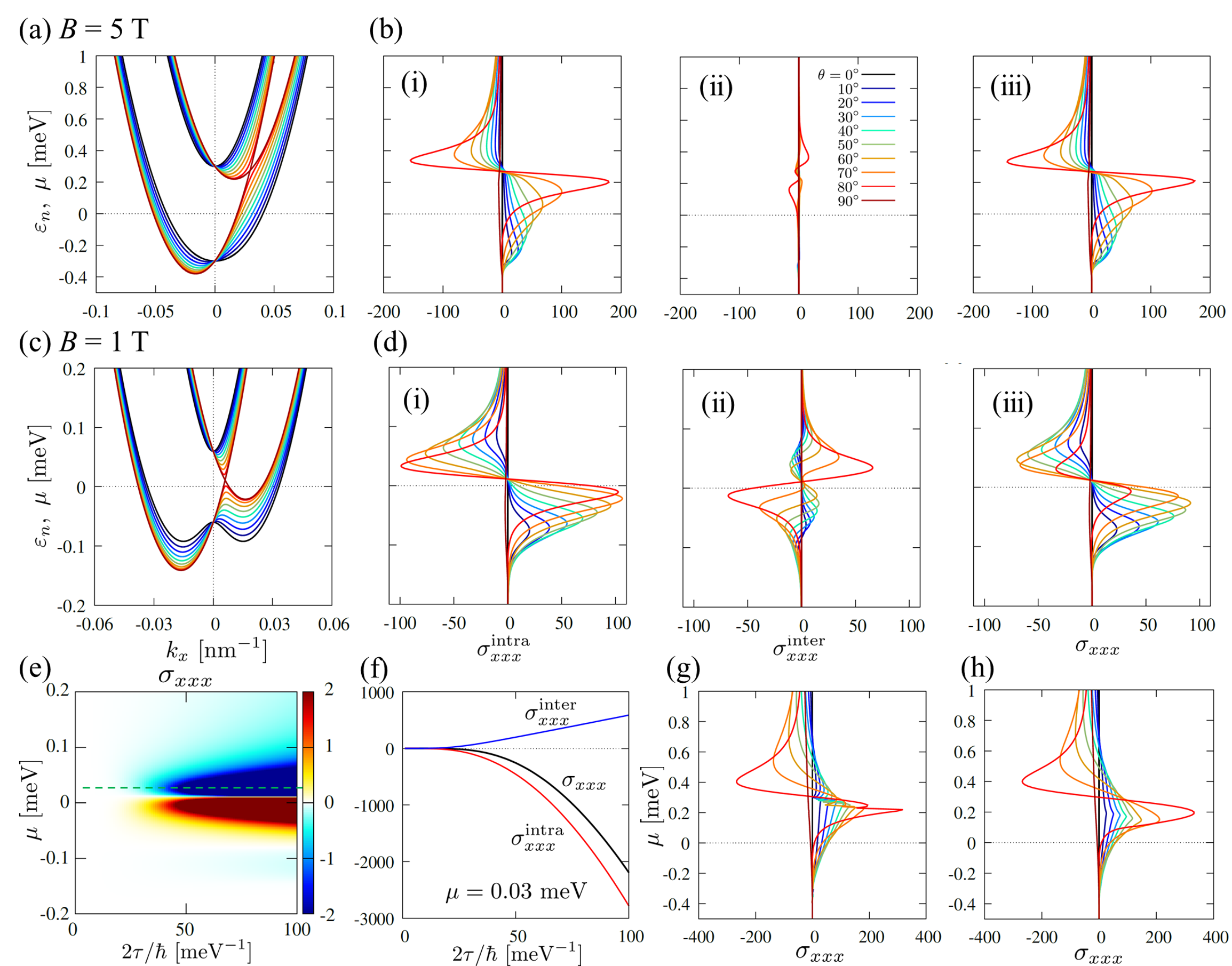}
\caption{
Nonlinear transport properties in 1D system. 
(a) Band structure in presence of a magnetic field of strength $B=5~{\rm T}$ applied in the $yz$ plane perpendicular to the nanowire axis along $x$ direction and rotated by the angle $\theta$ away from the $z$ axis. 
(b) Chemical potential $\mu$ counted from the SOI energy, i.e. level crossing at $\k=0$ for $\B=0$, and angle $\theta$ dependences at $B = 5~{\rm T}$ of (i) intraband contribution $\sigma_{xxx}^{\rm intra}$, (ii) interband contribution $\sigma_{xxx}^{\rm inter}$, and (iii) the sum of all contributions $\sigma_{xxx}$. 
Here, (c) and (d) are similar plots but for $B = 1~{\rm T}$, with same notations as in (a) and (b). 
(e) Relaxation time $\tau$ and $\mu$ dependence of $\sigma_{xxx}$, $\sigma_{xxx}^{\rm intra}$, and $\sigma_{xxx}^{\rm inter}$ at $\theta = 90^\circ$. (f) Plot of $\tau$-dependence of $\sigma_{xxx}$, $\sigma_{xxx}^{\rm intra}$, and $\sigma_{xxx}^{\rm inter}$ at $\mu = 0.03~{\rm meV}$ for $B = 1~{\rm T}$ and $\theta = 90^\circ$. The constant self-energy $\Sigma = i\gamma = i \hbar/(2\tau) = 0.033 i~{\rm meV}$ is employed for (b), (d), (e), and (f). [(g) and (h)] $\sigma_{xxx}$ considering the self-energy within the (g) Born approximation and (h) self-consistent Born approximation for $B=5~{\rm T}$. 
}
 \label{fig:1D}
\end{figure*}

\section{1D nanowire}
\label{sec:1D}

\begin{figure*}[t]
\includegraphics[width=180mm]{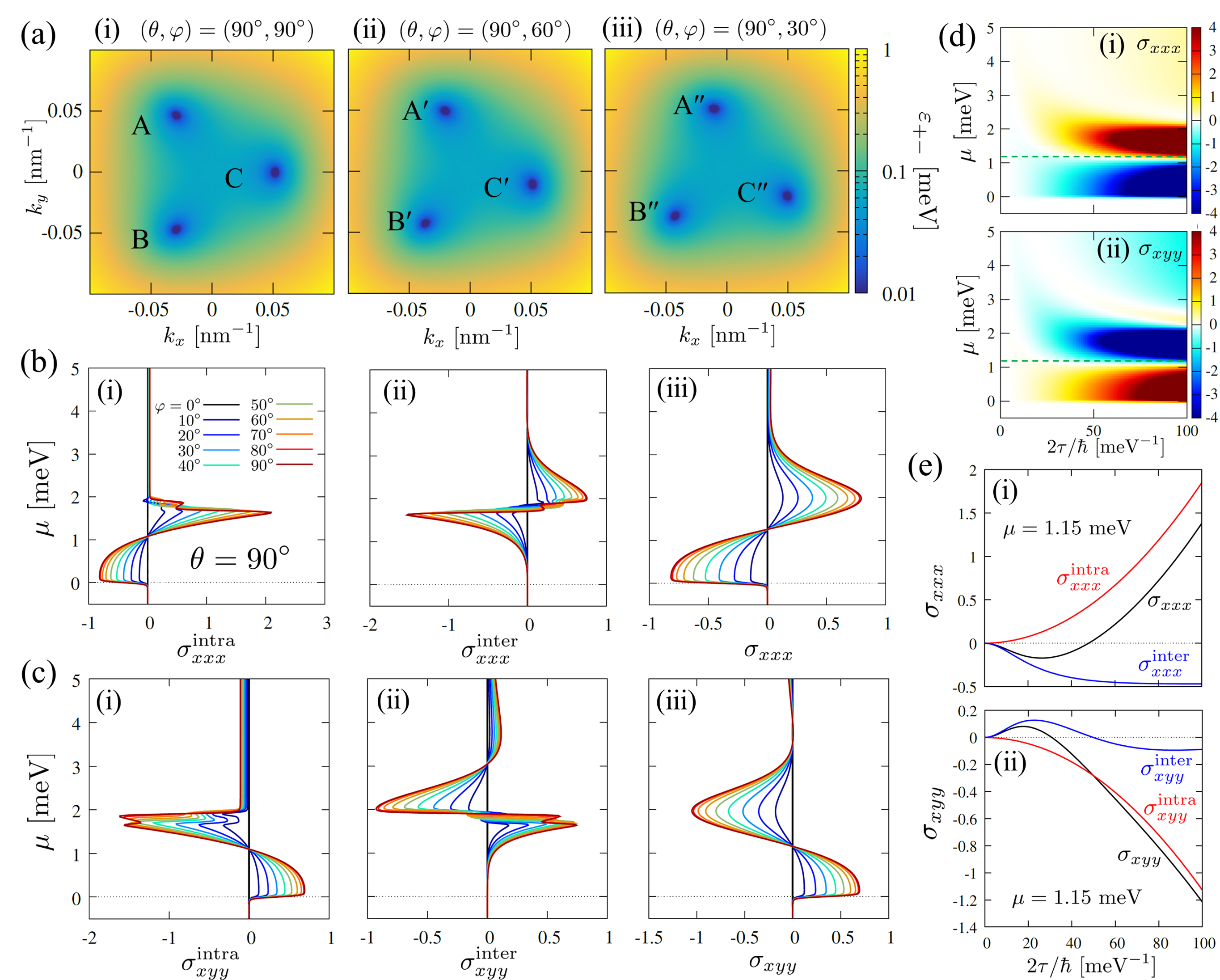}
\caption{ 
Nonlinear transport properties in a 2DHG without linear term of spin-orbit coupling, which simulates Ge [100] in the presence of an in-plane magnetic field.
(a) Difference of the band energy $\ve_{+-} \simeq \ve_{+} – \ve_{-}$ for several magnetic field angles; (i) $(\theta, \varphi) = (90^\circ, \ 90^\circ)$, (ii) $(90^\circ, \ 60^\circ)$, and (iii) $(90^\circ, \ 30^\circ)$. A, B, C, and their primed characters indicate nodal points.
(b) Chemical potential $\mu$ and the in-plane magnetic field angle $\varphi$ dependence of $\sigma_{xxx}$. (i) Intraband term, (ii) interband term, and (iii) their sum are plotted in each panel. 
(c) Chemical potential $\mu$ and the magnetic field angle $\varphi$ dependence of $\sigma_{xyy}$. Notations are the same to that of (b). 
(d) Relaxation time $\tau$ and chemical potential $\mu$ dependence of (i) $\sigma_{xxx}$ and (ii) $\sigma_{xyy}$. (e) $\tau$ dependence of the intraband, interband, and total (i) $\sigma_{xxx}$ and (ii) $\sigma_{xyy}$ at $\mu = 1.15$~$\rm meV$, represented by green dashed lines in (c). A constant self-energy $\Sigma^{\rR} = i\gamma = i \hbar/(2\tau ) = -0.033 i$~meV is used for (b)-(e). 
} 
 \label{fig:2DGe100} 
\end{figure*}

To gain insight into the different terms in Eq.~\eqref{eq:DC} we start by applying them to the simple situation of a quasi-one-dimensional nanowire (1DNW) with Hamiltonian~\cite{DLBLK} 
\begin{align}
\hat{H}_{k_x} = \frac{(\hbar k_x)^2}{2 m^*} - \a \sigma_y k_x - \b \sigma_y k_x^3 + {\bm \Delta} \cdot {\bm \sigma}, 
\end{align} 
where the $\a$ and $\b$ are the coefficients of the linear and cubic SOI with respect to the momentum $k_x$, $m^*$ is the effective mass, ${\bm \sigma} = (\s_x, \s_y, \s_z)$ are the Pauli matrices, and ${\bm \Delta} = \frac{1}{2} \mu_{\rm B} (g_x B_x, g_y B_y, g_z B_z)$ is a Zeeman field with the Bohr magneton $\mu_{\rm B}$, the effective $g$-factor $g_i$, and the magnetic field $B_i$. In this work we ignore orbital magnetic effects, leaving these for future studies. For the numerical estimates, we employ the values for Ge nanowires used in Ref.~\citenum{DLBLK}, namely $\hbar^2/(2m^* ) = 310~{\rm meV \ nm^2}$, $\a = 10~{\rm meV \ nm}$, $\b = 116~{\rm meV \ nm^3}$, and $g_i = 2.07$. The direction of the magnetic field is changed within the $yz$-plane, which is perpendicular to the nanowire. The spectrum is shown in Figs.~\ref{fig:1D}(a) and (c) under the magnetic field strength of $B=5~\text{T}$ and $B=1~\text{T}$, respectively. Nodal points occur in case of $\theta = 90^\circ$, and  gap openings are observed when we tilt the magnetic field toward the $z$ direction. 

Figs.~\ref{fig:1D}(b) and (d) show the second-order nonlinear conductivity $\sigma_{xxx}$ obtained by approximating the self-energy as an imaginary constant $\Sigma = -i\gamma = -0.033 i~{\rm meV}$ which corresponds to $\tau \sim 10~{\rm ps}$. $\sigma_{xxx}$ is calculated with the magnetic field strengths of (b) $B=5~\text{T}$ and (d) $B=1~\text{T}$, and is decomposed into the (i) intraband and (ii) interband contributions $\sigma_{xxx}^{\rm intra}$ and $\sigma_{xxx}^{\rm inter}$, respectively, as in Eqs.~\eqref{eq:decRRA} and \eqref{eq:decRA}. The term $\sigma_{xxx}$ has been calculated previously for such a system using $B=5~\text{T}$~\cite{DLBLK} and within the framework of the Boltzmann transport theory. Let us first focus on the intraband contribution. The maximal value is found around the band crossing energy $\mu \sim 0.3~{\rm meV}$ and $\mu \sim 0.02~{\rm meV}$ under $B=5~{\rm T}$ and $1~{\rm T}$, respectively. The peak is small when the magnetic field is oriented along the $y$ direction ($\theta = 90^\circ$), but it becomes approximately 15 times larger in case of  $\theta = 80^\circ$. Upon further increasing the out-of-plane $z$ component of the magnetic field, the maximal value of $\sigma_{xxx}$ decreases, reaching zero for $\theta = 0^\circ$ because the spin-up and spin-down branches of the energy spectrum have equal and opposite curvature. Meanwhile, $\sigma_{xxx}^{\text{inter}}$ is zero for $\theta = 90^\circ$ due to the symmetry of matrix elements while has finite value for $\theta < 80^\circ$ and gradually decreases as it approaches the $z$ direction, eventually reaching zero again. Importantly, reducing the magnetic field causes $\sigma_{xxx}^{\text{inter}}$ to have larger values, significantly changing the angular dependence of $\sigma_{xxx}$. In fact, the magnetic field angle dependence  of $\sigma_{xxx}$ is largely modified in case of $B=1~{\rm T}$, indicating the importance of the interband contribution for weaker fields. 

Next, we discuss the dependence on the relaxation time $\tau$. In Fig.~\ref{fig:1D}(c) we show the dependence of $\sigma_{xxx}$ on $\mu$ and $\tau$ for $\theta = 90^\circ$, indicating an increase in $\sigma_{xxx}$ around the energy near the band crossing points. Figure~\ref{fig:1D}(d) shows the intraband and interband contributions with total $\sigma_{xxx}$ for $\mu = 0.03~\text{meV}$. If $\tau$ is relatively small, the $\tau$-dependence of $\sigma_{xxx}$ is modified from $\tau^2$ due to the influence of $\sigma_{xxx}^{\text{inter}}$. On the other hand, when $\tau$ is sufficiently large, $\sigma_{xxx}$ is well approximated by $\sigma_{xxx}^{\text{intra}}$.

Figs.~\ref{fig:1D}(e) and (f) display the results obtained within the framework of Born approximation and the SCBA, respectively. We used $n_{\rm i}u^2 / V = 0.1~{\rm meV^2}$ for both calculations. In comparison to Fig.~\ref{fig:1D}(a), for the simple Born approximation, there is not much change in the overall $\mu$ dependence, but the peak positions are shifted. Moreover, dips are observed around the van Hove singularity since the self-energy is basically proportional to the density of states. However, after the self-consistent treatment with respect to the scattering process, these dips are suppressed since the van Hove singularities are smeared out. One more important inclusion are the off-diagonal components of the self-energy $\Sigma_{\uparrow \downarrow}$ and $\Sigma_{\downarrow \uparrow}$, which correspond to spin-flip processes~\cite{Taskin2017} and change the energy dependence of the nonlinear conductivity. For instance, the energy at which the sign changes is different between the diagonal self-energy case Fig.~\ref{fig:1D}(a) and after the inclusion of off-diagonal components Figs.~\ref{fig:1D}(e) and (f). 
Here, for simplicity, we employed the approximation of  an isotropic $g$-factor in the calculation, i.e. the Zeeman energy is constant. However, in general, the $g$-factor is anisotropic, such that $g_{x/y} \ll g_z$ and the Zeeman energy will not be constant when the magnetic field of constant strength is rotated.

\section{2DHG}
\label{sec:2D}

\begin{figure*}[t]
\includegraphics[width=180mm]{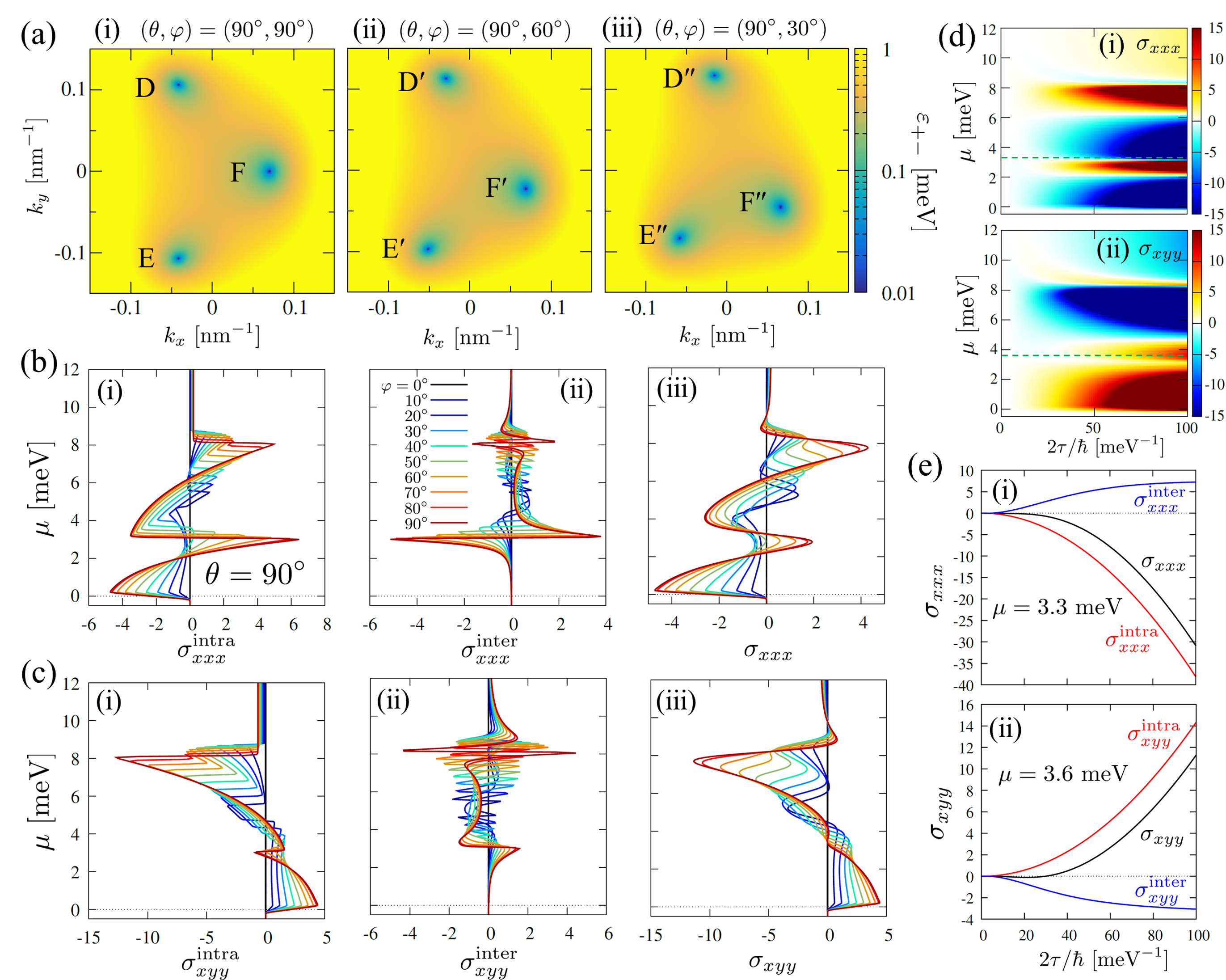}
\caption{ 
 Nonlinear transport properties of a 2DHG in Ge [110], i.e. with linear term of spin-orbit coupling. Figure order is identical to Fig.~\ref{fig:2DGe100}. 
}
 \label{fig:2DGe110}
\end{figure*}

Next we discuss the nonlinear transport properties in a 2DHG with both  linear and cubic SOI terms, described by the Hamiltonian~\cite{DLBLK}
\begin{widetext}
\begin{align}
\hat{H}_\k 
= \left(
\begin{array}{cc}
\frac{(\hbar k)^2}{ 2m^* } + \Delta_z & ik_- ( \a + \b_1 k_-^2 + \b_2 k_+^2 ) + \Delta_x - i\Delta_y \\
-ik_+ ( \a + \b_1 k_+^2 + \b_2 k_-^2 ) + \Delta_x + i\Delta_y & \frac{(\hbar k)^2}{ 2m^* } - \Delta_z \\
\end{array}
\right), 
\end{align}
\end{widetext}
where $k_\pm = k_x \pm ik_y$, and $\a$ and $\b_i$ are the strength of linear and cubic SOIs, respectively. We will primarily be interest in the case of an in-plane magnetic field and so, for simplicity, do not include orbital magnetic effects.

We also note that  for a 2DHG the treatment of the self-energy beyond the energy-constant approximation  is not necessary, since the density of states in a 2DHG is almost constant in energy. As such, the energy dependence of the relaxation time is also almost constant and only affects the nonlinear transport properties around the band bottom where the density of states does change considerably.

\subsection{Ge [100]}

The results presented in Fig.~\ref{fig:2DGe100} correspond to a model with material parameters associated with Ge [100]: ($\hbar^2/(2m^* ) = 620~{\rm meV \ nm^2}$, $\a=0$, $\b_1 = 190~{\rm meV \cdot nm^3}$, $\b_2 = 23.75~{\rm meV \cdot nm^3}$, $g_i = 0.207$) under the strength of the magnetic field $B = 5~{\rm T}$, similar to Ref.~\citenum{DLBLK}.

Fig.~\ref{fig:2DGe100}(a) shows the energy difference between two bands $\pm$ at each momentum, $\ve_{+-} \equiv \ve_{+} – \ve_{-}$, for three different in-plane magnetic field directions: (i) $(\theta, \varphi) = (90^\circ, \ 90^\circ)$, (ii) $(90^\circ, \ 60^\circ)$, and (iii) $(90^\circ, \ 30^\circ)$. Since we restricted the magnetic field direction to the $xy$ plane, the gap remains closed and the position of the nodal point in the momentum space depends on the angle of the magnetic field. In case of $(\theta, \varphi) = (90^\circ, \ 90^\circ)$, the energy of nodal points A, B, and C are $E_{\rm A} = E_{\rm B} = 1.91$~$\rm meV$ and $E_{\rm C} = 1.67$~$\rm meV$, respectively, namely A and B have the same energy. Besides, as for $(\theta, \varphi) = (90^\circ, \ 60^\circ)$, the energy at A$'$, B$'$, and C$'$ are $E_{\rm A'} =1.97$~$\rm meV$, $E_{\rm B'} = 1.80$~$\rm meV$ and $E_{\rm C'} = 1.71$~$\rm meV$, and for $(\theta, \varphi) = (90^\circ, \ 30^\circ)$, A$''$, B$''$, and C$''$ have the energies of $E_{\rm A''} = 1.97$~$\rm meV$, $E_{\rm B''} = 1.71$~$\rm meV$ and $E_{\rm C''} = 1.80$~$\rm meV$, respectively, which means that each nodal point is located at a mutually different energy. In addition, $E_{\rm B'} = E_{\rm C''}$ and $E_{\rm C'} = E_{\rm B''}$ are satisfied because of the symmetry. We note that the difference of the energy at the nodal point is $\sim 0.3$~${\rm meV}$, which is comparatively small. This value should be compared with Ge [110] system which we discuss later, and important for understanding the behavior of the nonlinear transport properties. 

Next, Figs.~\ref{fig:2DGe100}(b) and (c) show the chemical potential and in-plane magnetic field angle $\varphi$ dependence of $\sigma_{xxx}$ and $\sigma_{xyy}$, respectively. $\sigma_{xyy}$ under the in-plane field represents a planar nonlinear Hall effect~\cite{Huang2023}. We assume the energy independent relaxation time which is assumed to be $\tau \sim 10 \ {\rm ps}$, similar to 1DNW case. Focusing on the intraband component $\sigma_{xii}^{\rm intra}$, peaks around the nodal points are observed that when the magnetic field is in the plane ($\theta=90^\circ$), as mentioned in the previous Boltzmann analysis~\cite{DLBLK}. As we change the magnetic field angle, we should observe three sharp peaks corresponding to the nodal points which we discussed in previous paragraph, however, it is not appearing clearly in the plots because the energy difference of two upper nodal points is too small. For $\mu \gtrsim 2~{\rm meV}$, the values are largely suppressed and energy-independent. In contrast, for $\sigma^{\rm inter}$, the substantial contribution is found around the nodal point, and the spikes observed in both $\sigma_{xxx}^{\rm intra}$ and $\sigma_{xyy}^{\rm intra}$ are smeared out as shown in Fig.~\ref{fig:2DGe100}(b) (iii) and (c) (iii). Moreover, $\sigma_{xii}^{\rm inter}$ have the considerable value even in $\mu \gtrsim 2~{\rm meV}$ where the intraband components are suppressed. Note that the Berry curvature dipole term Eq.~\eqref{eq:BCD} contained in $\sigma_{xyy}^{\rm inter}$ is absent since here we restrict the direction of the magnetic field to the in-plane direction. Hence, we emphasize here that the band geometry term $\sigma_{xii}^{\rm QG}$ plays an important role in the present second-order responses. This will be the case also for Ge [110] 2DHG which we discuss later. This large modification caused by the visible interband terms is due to the comparatively small band energy difference. Especially, the energy difference around the nodal points are remarkably small, which enhances the interband term. This should be compared with the case of Ge [110] which we discuss in the next subsection. 

Furthermore, in Figs.~\ref{fig:2DGe100}(d) and (e), typical dependences on the relaxation time $\tau$ are shown. The $\tau$ dependence, considering only intraband terms, should be $\tau^2$ for all values of $\mu$, however, as shown in Fig.~\ref{fig:2DGe100}(d), we observed an anomalous $\tau$ dependence which largely deviates from a $\tau^2$ behavior, especially around $\mu \sim 1.3$ and $\mu \gtrsim 2~{\rm meV}$. Fig.~\ref{fig:2DGe100}(e) show the plots of $\sigma_{xii}^{\rm intra}$, $\sigma_{xii}^{\rm inter}$, and their sum at $\mu = 1.15~{\rm meV}$, highlighting the anomalous $\tau$ dependence due to the interband contribution. In particular, we observe an approximate $\tau^2$ dependence in the region where $\tau$ is small, which is consistent with Eq.~(\ref{eq:domQM}). These results demonstrates that in the disordered systems, with small $\tau$, the presence of interband terms significantly alters the nonlinear transport characteristics, even in Ge [100]. On the other hand, in cleaner systems with large $\tau$ the interband contribution has lower order, $\sim \tau^0$, compared to the intraband contribution, which still scales as $\sim \tau^2$. Therefore, it is desirable to investigate clean systems with larger $\tau$ in order to observe these nodal structure and determine spin-obit interaction~\cite{DLBLK}.
  
\subsection{Ge [110]}

Next, we consider the Ge [110] growth direction, as shown in Fig.~\ref{fig:2DGe110}. In particular, in Fig.~\ref{fig:2DGe110} we present equivalent results to those shown in Fig.~\ref{fig:2DGe100}, but for parameters corresponding to Ge [110] ($\hbar^2/(2m^* ) = 620~{\rm meV \ nm^2}$, $\a=1.5~{\rm meV \cdot nm}$, $\b_1 = 190~{\rm meV \cdot nm^3}$, $\b_2 = 23.75~{\rm meV \cdot nm^3}$, $g_i = 1.244$) under the strength of the field $B = 5~{\rm T}$, as also used in the Ge [100] case. 

In Fig.~\ref{fig:2DGe110}(a), the energy difference $\ve_{+-}$ exhibits similar behavior to Ge [100]; preserving nodal points for an in-plane magnetic field.  However, due to the existence of the linear SOI and large $g$-factor, $\ve_{+-}$ is, in general, significantly larger than Ge [100]. This leads to the suppression of the interband terms, which we will see later. The energy of nodal points for $(\theta, \varphi) = (90^\circ, \ 90^\circ)$ are $E_{\rm D} = E_{\rm E} = 8.12$~$\rm meV$ and $E_{\rm F} = 3.06$~$\rm meV$, respectively, D and E have same energy because of symmetry. As for $(\theta, \varphi) = (90^\circ, \ 60^\circ)$, their energies are $E_{\rm D'} = 7.43$~$\rm meV$, $E_{\rm E'} = 8.51$~$\rm meV$ and $E_{\rm F'} = 3.29$~$\rm meV$, and for $(\theta, \varphi) = (90^\circ, \ 30^\circ)$, $E_{\rm D''} = 6.37$~$\rm meV$, $E_{\rm E''} = 8.70$~$\rm meV$ and $E_{\rm F''} = 3.91$~$\rm meV$, respectively, again the energies of each nodal point are different. In addition, this time we observe $E_{\rm E'} \neq E_{\rm F''}$ and $E_{\rm F'} \neq E_{\rm E''}$ as a result of the presence of linear SOI. The difference between the energies of nodal points are few $\rm meV$ in Ge [110], which is considerably larger than Ge [100]. 

Figure~\ref{fig:2DGe110}(b) and (c) show $\sigma_{xxx}$ and $\sigma_{xyy}$ respectively. As we change the magnetic field angle in the $xy$-plane, the presence of band crossing points leads to a significant interband contribution and, as a result, the peaks in the intraband component are smeared out  due to the interband (band geometry) contributions,  similar to the Ge [100] case. The significant difference between Ge [100] and Ge [110] is that the energies of the three band crossing points are substantially different when the in-plane magnetic field is rotated in the $xy$-plane. Consequently, even though the interband contribution can average out the peaks in the intraband component, the overall energy dependence remains unchanged. For instance, in Fig.~\ref{fig:2DGe110}(c) for the case of $(\theta, \ \varphi)= (90^\circ, \ 60^\circ)$, it shows that the peak observed in $\sigma_{xxx}^{\rm intra}$ persists in $\sigma_{xxx}$, demonstrating its resilience to suppression from the interband component. This angular dependence of in-plane magnetic fields can therefore also serve as a new fingerprint of the presence of linear SOI. In addition, Figs.~\ref{fig:2DGe100}(d) and (e) comprehensively shows the $\tau$ dependence of $\sigma_{xxx}$ and $\sigma_{xyy}$, which have quite similar properties to the Ge [100] case. As previously, the $\tau$ dependence deviates from what Boltzmann formula predicts, due to the interband terms. 

\section{Discussion}
\label{sec:discussion}

\subsection{Detecting SOI} \label{subsec:soi} 

We have investigated how interband contributions modify nonlinear transport coefficients, going beyond simple semiclassic approximations. We found that the interband contributions are largest when the energy difference between the bands, $\ve_{+-}$, is small but nonzero. As a result, according to our findings, there are two optimal scenarios where the interband terms are suppressed: i) in a clean system with weak disorder and therefore large $\tau$, and/or ii) in the presence of a large magnetic field resulting in a large $\varepsilon_{nm}$. The first point is based on the fact that the intraband term is proportional to $\tau^2$ and interband contributions are lower order in $\tau$ (higher order of $\gamma$) in clean systems. As such, in a sufficiently clean system with a large $\tau$, the intraband effect will exceed the interband contribution. The second point is based on the fact that the interband term is more pronounced when the difference in band energies is small. 

To summarize, the criterion for neglecting the interband terms is determined by the relationship between the band energy difference $\ve_{+-}$ and the damping rate $\gamma=\frac{\hbar}{2\tau}$, such that it is important to make the region which satisfies $\ve_{+-} / \gamma \gg 1$ as large as possible. Given that the mean free path in Ge can reach several 10s of $\mu$m, this should be satisfied in clean Ge samples \cite{Stehouwer23}. Note, however, that the interband contribution is important, even in a clean system, if the chemical potential is located sufficiently close to a nodal point (see Appendix~\ref{sec:taudep}). 

\subsection{Vertex corrections} 

We  emphasize that our analysis neglects the contribution from vertex corrections arising from impurity scattering. Although we do not expect such corrections to strongly alter the qualitative results in the cases considered here, it should be noted that, for instance, previous studies have shown that side-jump and skew-scattering process can give  considerable contributions in the general two band system with finite Berry curvature dipole~\cite{DWSLX}. Furthermore, since our formalism can treat the effects of impurity scattering in a fully microscopic way it should be possible to implement vertex corrections in a gauge invariant way via Ward identities. The implementation of vertex corrections could, therefore, be of considerable interest for future studies, especially for systems with a finite Berry curvature dipole.

\section{Summary}
\label{sec:summary}

We have developed a microscopic formulation to analyze nonlinear transport including both intraband and interband contributions in the presence of a magnetic field inducing Zeeman splittings (but without orbital effects). First, we derived a microscopic formulation to calculate nonlinear transport coefficients using the Keldysh formalism and confirmed its consistency with the results of previous studies based on the Boltzmann equation approach. Then, we defined the intraband and interband terms using the band representation and investigated the dependence of each term on the relaxation time at zero temperature, confirming that the interband term is important when the band energy difference is comparable to (or smaller than) the damping rate. In particular, within this framework, we find that the interband contributions arise due to the band (quantum) geometry and Berry curvature dipole. As a test for this formulation, we considered transport in a 1DNW and found that interband terms are the largest close to the nodal point and in the case of weak magnetic fields. Furthermore, we found that a full treatment of the self-energy by Born approximation can significantly alter the nonlinear transport properties in a 1DNW. As a second example, we considered the 2DHG and, in particular, models corresponding to Ge [100] and Ge [110]. In both cases, the interband term is found to increase around the nodal point, similar to what was observed for the 1DNW. The magnetic field angle dependences in the $xy$-plane are significantly different between Ge [100] and Ge [110], reflecting the nature of SOI. Our results provide a microscopic framework to calculate nonlinear transport coefficients beyond standard Boltzmann descriptions. In addition, we found that a simple interpretation of SOI contributions to nonlinear transport signatures in semiconductor nanostructures necessitates the use of clean samples, where interband terms can be neglected.

\begin{acknowledgements}
We thank Y. Michishita and T. Yamaguchi for the fruitful discussions. K.N. is supported by JSPS KAKENHI Grant Number JP21K13875.  
This work was supported as a part of NCCR SPIN, a National Centre of Competence in Research, funded by the Swiss National Science Foundation (grant number 225153).
\end{acknowledgements}

\appendix

\begin{widetext}

\section{Derivation of Keldysh expressions up to second order in external fields}
\label{sec:general_detail}

\subsection{General}

We give the action $S = S_0 + S'$ to describe the model: 
\begin{align}
S_0 
&= \int dt dt' \int d\r d\r' \, 
\bpsi_n^a (\r,t) [ G_{nm}^{-1} (\r-\r',t-t') ]^{ab}  \psi_n^b ( \r',t )  , 
\\
S' 
&= - \int dt \int d\r \ \bpsi_n^c (\r,t) X_{i,nm} (\r,t) \psi_m^c (\r,t) F_i (\r,t) , 
\end{align}
whose Fourier expressions are
\begin{align}
S_0 &= \int \frac{d\ve}{2\pi} \sum_\k \bpsi_{n,\k}^a (\ve) [ G_{\k,nm}^{-1} (\ve) ]^{ab} \psi_{m,\k}^b, 
\\
S' 
&= - \int \frac{d\w}{2\pi} \sum_{\k, \q} \bpsi_{n,\k+\q/2}^c (\w) X_{i,nm} (\k,\ve;-\q,-\w) \psi_{m,\k-\q/2}^c (\w) F_{i,\q} (\w), 
\end{align}
\end{widetext}
where $\bpsi$ and $\psi$ are the Grassmann numbers, $\bm F$ and $\bm X$ are the external field and coupled physical quantity, and $a$, $b$, and $c$  specify {\it forward} and/or {\it backward} time path defined on {\it closed time contour}. Using Gaussian integral, we obtain the Green function,
\begin{align}
G_{nm}^{a b} (\k,\k',\ve,\ve') &\equiv -i\langle \psi_{n\k}^a (\ve) \bpsi_{m\k'}^b (\ve') \rangle_{\rm full} 
\nonumber \\ 
&= -i \int {\cal D} [\bpsi \psi]  \psi_{n\k}^a (\ve) \bpsi_{m\k'}^b (\ve') \expo^{i(S_0 + S')} .
\label{eq:fullgf}
\end{align} 
After the Keldysh rotation, the superscripts denote the components of the Keldysh space~\cite{Kamenev}. The expectation value of a physical observable $\hat{O}$ can be calculated from 
\begin{align}
\langle \hat{O} \rangle = -\frac{i}{2} {\rm tr} [ \hat{O} \hat{G}^{\rm K} ]
\label{eq:O}
\end{align}
where $\hat{G}^\rK$ is the Keldysh Green function, and $\rm tr$ is the trace taken over all  quantum numbers (e.g., momentum, frequency, spin, sublattice, orbital, etc.). We concentrate here on the Keldysh component. 

We can consider expanding the Green function Eq. (\ref{eq:fullgf}) with respect to the coupling to the external field $S_{\rm ext}$. Below we use four-momentum vectors $k = (\k,\ve)$ and $q = (\q,\w)$ which specify the momentum of the Green function and the external field, respectively, and employ the notations $k_\pm = k \pm q/2$ and $\ve_\pm = \ve \pm \w/2$, as well as $k_{\pm\pm} = k \pm q/2 \pm q'/2$, $k_{0\pm} = k \pm q'/2$, $k_{\pm 0} = k \pm q/2$, and $\ve_{\pm\pm} = \ve \pm \w/2 \pm \w'/2$. Performing a perturbative expansion in the external fields and using Wick's theorem, the 0th, 1st, and 2nd order terms of the Keldysh Green function are expressed as  

\begin{widetext}
\begin{align}
G_{nm}^{\rm K(0)} (k) 
&= G_{nm}^{\rm K(0)} (k,k') \delta (k-k')
= -i\langle \psi_{n}^1 (k) \bpsi_{m}^1 (k) \rangle_0 
= (1 - 2f(\ve) ) ( G^\rR (k) - G^\rA (k) )_{nm} ,
\label{eq:K0}
\\
G_{nm}^{\rm K(1)} (k;q)
&=-\int {\cal D} [\bpsi \psi] \psi_{n}^a (k) \bpsi_{m}^b (k') \expo^{iS_0} \int \frac{d\ve''}{2\pi} 
      \sum_{\k''} \bpsi_{n'}^\g (k''_+) X_{i,n'm'} (k'';-q) \psi_{m'}^\g (k''_-) F_i (q)  \nonumber 
\\
&= -\int \frac{d\ve''}{2\pi} \sum_{\k''} \left[  \hat{G} (k_+) \hat{O}_i (k;-q) \hat{G} (k_-) \right]_{nm}^{\rm K} F_i (q) 
 \nonumber 
\\
&= - 2 \Bigl[ \left\{ f(\ve_+) - f(\ve_-) \right\} G_{nn'}^\rR (k_+) X_{i,n'm'} (k;-q) G_{m'm}^\rA (k_-) \nonumber 
\\
&\quad +  f(\ve_-) G_{nn'}^\rR (k_+) X_{i,n'm'} (k;-q) G_{m'm}^\rR (k_-) 
- f(\ve_+) G_{nn'}^\rA (k_+) X_{i,n'm'} (k;-q) G_{m'm}^\rA (k_-) \Bigr] F_i (q) ,
\label{eq:K1}
\\
G_{nm}^{\rm K(2)} (k;q,q')
&=-\frac{1}{2} \int {\cal D} [\bpsi \psi] \psi_{n}^a (k) \bpsi_{m}^b (k') \expo^{iS_0}
\nonumber \\
& \quad \times \int \frac{d\ve'' d\ve'''}{(2\pi)^2} 
      \sum_{\k'',\k'''} \bpsi_{n'}^c (k''_+) X_{i,n'm'} (k'';-q) \psi_{m'}^c (k''_-) 
     				   \bpsi_{n''}^d (k'''_+) X_{j,n''m''} (k''';-q') \psi_{m'}^d (k'''_-) F_i (q) F_j (q') 
\nonumber \\ 
& \quad + \{ (i,q) \leftrightarrow (j,q') \}
\nonumber \\
&= -\frac{1}{2}
\Bigl[ \hat{G} (k_{++}) \hat{X}_i (k_{0+};-q) \hat{G} (k_{-+}) \hat{X}_j (k_{-0};-q') \hat{G} (k_{--})   
\nonumber \\
&\quad \quad + \hat{G} (k_{++}) \hat{X}_j (k_{+0};-q') \hat{G} (k_{+-}) \hat{X}_i (k_{0-};-q) \hat{G} (k_{--}) 
\Bigr]_{nm}^{\rm K} F_i (q) F_j (q') 
\nonumber \\
&=-\Bigl[
\left\{ f(\ve_{++}) - f(\ve_{-+}) \right\} \hat{G}^\rR (k_{++}) \hat{X}_i (k_{0+};-q) \hat{G}^\rA (k_{-+}) \hat{X}_j (k_{-0};-q') \hat{G}^\rA (k_{--})  
\nonumber \\
&\quad + \left\{ f(\ve_{-+}) - f(\ve_{--}) \right\} \hat{G}^\rR (k_{++}) \hat{X}_i (k_{0+};-q) \hat{G}^\rR (k_{-+}) \hat{X}_j (k_{-0};-q') \hat{G}^\rA (k_{--})
\nonumber \\
&\quad + f(\ve_{++}) \left\{ \hat{G}^\rA (k_{++}) \hat{X}_i (k_{0+};-q) \hat{G}^\rA (k_{-+}) \hat{X}_j (k_{-0};-q') \hat{G}^\rA (k_{--}) \right\} 
\nonumber \\
&\quad - f(\ve_{--}) \left\{ \hat{G}^\rR (k_{++}) \hat{X}_i (k_{0+};-q) \hat{G}^\rR (k_{-+}) \hat{X}_j (k_{-0};-q') \hat{G}^\rR (k_{--}) \right\} 
\nonumber \\
&\quad + \{ (i,q) \leftrightarrow (j,q') \}  \Bigr]_{nm}  F_i (q) F_j (q') .
\label{eq:K2}
\end{align}
Note that these representations do not depend on the choice of the basis. We will move to the band representation later, which is useful for the analysis of the effective model. 

\subsection{Nonlinear optical conductivity}

Let us calculate the nonlinear optical conductivity and check the consistency with  previous studies~\cite{PMOM,JL,MP}. Let us omit the obvious superscripts and subscripts hereafter. After the minimal substitution, we substitute $\hat{O} \simeq - e \hat{\cV}_\a + e^2 A_\b \hat{\cV}_{\a\b} - (e^3/2) A_\b A_\g \hat{\cV}_{\a\b\g}$ and $\hat{X}_i \Phi_i \simeq - eA_\a \hat{\cV}_\a + (e^2/2) A_\a A_\b \hat{\cV}_{\a\b}$ in Eqs.~(\ref{eq:O})-(\ref{eq:K2}) and taking the limit $\q,\q' \to { 0}$  to consider a uniform electric field, we obtain the second order AC current: 
\begin{align}
\langle j_\a^{(2)} \rangle (\w_1,\w_2) &= 
-\frac{ie^3 }{2\w_1 \w_2} E_\b (\w_1) E_\g (\w_2)  
\nonumber \\ 
&\quad \times {\rm tr} \Bigl[ f(\ve) \Bigl\{
\hat{\cV}_{\a\b\g} \GRmGA + \hat{\cV}_{\a\b} \hat{G}^\rR (\ve + \w_2) \hat{\cV}_\g \GRmGA + \hat{\cV}_{\a\b} \GRmGA \hat{\cV}_\g \hat{G}^\rA (\ve - \w_2) \nonumber \\
&\quad \quad \quad \quad \quad \quad \quad \quad \quad \quad \quad \quad \quad + \hat{\cV}_{\a\g} \hat{G}^\rR (\ve + \w_1) \hat{\cV}_\b \GRmGA + \hat{\cV}_{\a\g} \GRmGA \hat{\cV}_\b \hat{G}^\rA (\ve - \w_1)  \nonumber \\
&\quad \quad \quad \quad \quad \quad \quad \quad \quad \quad \quad \quad \quad + \hat{\cV}_{\a} \hat{G}^\rR (\ve + \Omega) \hat{\cV}_{\b\g} \GRmGA + \hat{\cV}_{\a} \GRmGA \hat{\cV}_{\b\g} \hat{G}^\rA (\ve - \Omega) \nonumber \\
&\quad \quad + \hat{\cV}_\a \hat{G}^\rR (\ve + \Omega) \hat{\cV}_\b \hat{G}^\rR (\ve + \w_2) \hat{\cV}_\g \GRmGA + \hat{\cV}_\a \hat{G}^\rR (\ve + \Omega) \hat{\cV}_\g \hat{G}^\rR (\ve + \w_1) \hat{\cV}_\b \GRmGA \nonumber \\
&\quad \quad + \hat{\cV}_\a \hat{G}^\rR (\ve + \w_1) \hat{\cV}_\b \GRmGA \hat{\cV}_\g \hat{G}^\rA (\ve - \w_2) + \hat{\cV}_\a \hat{G}^\rR (\ve + \w_2) \hat{\cV}_\g \GRmGA \hat{\cV}_\b \hat{G}^\rA (\ve - \w_1) \nonumber \\
&\quad \quad + \hat{\cV}_\a \GRmGA \hat{\cV}_\b \hat{G}^\rA (\ve - \w_1) \hat{\cV}_\g \hat{G}^\rA (\ve - \Omega) + \hat{\cV}_\a  \GRmGA \hat{\cV}_\g \hat{G}^\rA (\ve + \w_1) \hat{\cV}_\b  \hat{G}^\rA (\ve - \Omega) \Bigr\} \Bigr] \nonumber \\
&= \sigma_{\a\b\g} (\w_1,\w_2) E_\b (\w_1) E_\g (\w_2),
\label{eq:ja}
\end{align}
where $\Omega = \w_1 + \w_2$. We draw the corresponding diagrams in Fig.~\ref{fig:K2}. Thus, we derived the formula consistent with Ref.~\cite{JL,MP} (except for the global prefactor 1/2). 

\subsection{Validity check for DC limit}
\label{sec:DClimit}

Let us quickly check that the DC limit is uniquely determined. It is sufficient to show that the term proportional to $\w_i^n/\w_1 \w_2$ in $\sigma_{\a\b\g}^{(2)}$ is zero. Namely, if we write
\begin{align}
\sigma_{\a\b\g}^{(2)} (\w_1,\w_2) = \frac{ie^3}{2} {\rm tr} \Biggl[ \frac{A_0}{\w_1 \w_2} &+ \sum_{n \geq 1} \frac{A_n \w_1^n+ A'_n \w_2^n}{\w_1 \w_2} + ({\rm Regular \ part}) \Biggr],   
\end{align}
we need to show that ${\rm tr} A_n = {\rm tr} A'_n = 0$. For $n=0,1$ this was shown in Ref.~\cite{MP}. Generalizing this result, we now demonstrate that ${\rm tr} A_n$, ${\rm tr} A'_n $ vanish for general $n$. Indeed, collecting all  the terms proportional to $\omega_i^n$, we get 
\begin{align}
A'_n &= f(\ve) \frac{1}{n!} \left[ \hat{\cV}_{\a\b} ( \partial_\ve \hat{G}^\rR ) \hat{\cV}_\g \hat{G}^\rR + \hat{\cV}_\a \hat{G}^\rR \hat{\cV}_{\b\g} \hat{G}^\rR +  \sum_{\ell=0}^n {}_n C_\ell \hat{\cV}_\a ( \partial_\ve^{n-\ell} \hat{G}^\rR ) \hat{\cV}_\b (\partial_\ve^\ell \hat{G}^\rR ) \hat{\cV}_\g \hat{G}^\rR + \hat{\cV}_\a ( \partial_\ve^n \hat{G}^\rR ) \hat{\cV}_\g \hat{G}^\rR \hat{\cV}_\b \hat{G}^\rR \right] \nonumber \\
&- f(\ve) \frac{1}{n!} \left[ \hat{\cV}_{\a\b} ( \partial_\ve \hat{G}^\rA ) \hat{\cV}_\g \hat{G}^\rA + \hat{\cV}_\a \hat{G}^\rA \hat{\cV}_{\b\g} \hat{G}^\rA +  \sum_{\ell=0}^n  {}_n C_\ell \hat{\cV}_\a ( \partial_\ve^{n-\ell} \hat{G}^\rA ) \hat{\cV}_\b (\partial_\ve^\ell \hat{G}^\rA ) \hat{\cV}_\g \hat{G}^\rA + \hat{\cV}_\a ( \partial_\ve^n \hat{G}^\rA ) \hat{\cV}_\g \hat{G}^\rA \hat{\cV}_\b \hat{G}^\rA \right] \nonumber \\
&+ \frac{\partial_\ve f(\ve) }{(n-1)!} \Biggl[ \hat{\cV}_{\a\b} (\partial_\ve^{n-1} \hat{G}^\rR) \hat{\cV}_\g \hat{G}^\rA + \hat{\cV}_\a (\partial_\ve^{n-1} \hat{G}^\rR) \hat{\cV}_{\b\g} \hat{G}^\rA \nonumber \\ 
&\quad \quad \quad \quad \quad \quad \quad \quad \quad \quad \quad \quad \quad \quad \quad \quad 
+ \sum_{\ell=0}^{n-1} {}_{n-1} C_\ell \hat{\cV}_\a (\partial_\ve^{n-1-\ell} \hat{G}^\rR) \hat{\cV}_\b (\partial_\ve^\ell \hat{G}^\rR) \hat{\cV}_\g \hat{G}^\rA + \hat{\cV}_\a (\partial_\ve^{n-1} \hat{G}^\rR) \hat{\cV}_\g \hat{G}^\rA \hat{\cV}_\b \hat{G}^\rA \Biggr] \nonumber  \\
&= \frac{1}{n!} 
\partial_\b \left[ \hat{\cV}_\a (\partial_\ve^n \hat{G}^\rR ) \hat{\cV}_\g \hat{G}^\rR - {\rm c.c.} \right] 
+ 
\frac{1}{(n-1)!}\partial_\b \left[ \hat{\cV}_\a (\partial_\ve^{n-1} \hat{G}^\rR ) \hat{\cV}_\g \hat{G}^\rA \right] .
\end{align}
A similar transformation applies to $A_n$, therefore, again these terms reduce to surface terms and hence their trace vanishes. Thus, we checked that we can take the DC limit without any divergence or uncertainty as expected.

\section{Band representation}
\label{sec:bandrep}

In this appendix, we derive the band representation for the physical description. Let us put here $-e^3 / (2\pi) = 1$ to simplify the equations. First, we note that
\begin{align}
i\partial_\a ( \lan | \ram ) &= i \dalan | \ram + i \lan | \daram = [-i \lam | \daran ]^* - [- i \lan | \daram ] = (A_\a^{mn})^* - A_\a^{nm}  = 0,
\\
-i \partial_\b \partial_\a ( \lan | \ram ) &= -i\left[ \langle \partial_\a \partial_\b n | \ram + \dblan | \daram + \dalan | \dbram + \lan | \partial_\a \partial_\b m \rangle \right]
\nonumber \\
&= -i \partial_\a ( \lan | \dbram ) - i \partial_\b ( \dalan | \ram ) = \partial_\a A_\b^{nm} - \partial_\b (A_\a^{mn})^* = \partial_\a A_\b^{nm} - \partial_\b A_\a^{nm} =0,
\end{align}
from which we see the useful relations $A_i^{nm} = (A_i^{mn})^*$ and $\partial_\a A_\b^{nm} = \partial_\b A_\a^{nm}$. 
Using Eqs.~\eqref{eq:uni}, \eqref{eq:UAU}, and \eqref{eq:UABU}, the extrinsic terms $\sigma_{\a\b\g}^{\rm RRA}$ and $\sigma_{\a\b\g}^{\rm RA}$ are expressed as
\begin{align}
\sigma_{\a\b\g}^{\rm RRA} &\equiv {\rm Im} \ {\rm tr} [ \hat{\cV}_\a (\hat{G}^\rR)^2 \hat{\cV}_\b \hat{G}^\rR \hat{\cV}_\g \hat{G}^\rA + (\b \leftrightarrow \g) ]
\nonumber \\
&=
{\rm Im} \sum_{nm\ell} \sum_\k [ \lan | \hat{\cV}_\a | \ram (G_m^\rR)^2 \lam |\hat{\cV}_\b | \ral G_\ell^\rR \lal |\hat{\cV}_\g | \ran G_n^\rA + (\b \leftrightarrow \g) ] 
\nonumber \\
&= {\rm Im} \sum_{nm\ell} \sum_\k \biggl[
(\partial_\a \ve_n) (\partial_\b \ve_m) (\partial_\g \ve_\ell) \delta_{nm} \delta_{m\ell} \delta_{\ell n} -
\ve_{m\ell} \ve_{\ell n} (\partial_\a \ve_n) \delta_{nm} A_\b^{m\ell} A_\g^{\ell n} 
-
\ve_{nm} \ve_{\ell n} (\partial_\b \ve_m) A_\a^{nm} \delta_{m\ell} A_\g^{\ell n} 
\nonumber \\ &\quad \quad \quad \quad \quad -
\ve_{nm} \ve_{m\ell} (\partial_\g \ve_\ell) A_\a^{nm} A_\b^{m\ell} \delta_{\ell n} 
+
i\ve_{nm} \ve_{m\ell} \ve_{\ell n} A_\a^{nm} A_\b^{m\ell} A_\g^{\ell n}  + (\b \leftrightarrow \g) \biggr]
 (G_m^\rR)^2 G_\ell^\rR G_n^\rA
\label{eq:full1}
\\
&=
{\rm Im} \sum_\k \sum_n \biggl[ 2 v_\a^{(n)} v_\b^{(n)} v_\g^{(n)} (G_n^\rR)^3 G_n^\rA 
+ \sum_{\ell} \ve_{n\ell}^2 v_\a^{(n)} ( A_\b^{n\ell} A_\g^{\ell n} + A_\g^{n\ell} A_\b^{\ell n} ) G_\ell^\rR (G_n^\rR)^2 G_n^\rA \nonumber \\
&\quad \quad \quad + \sum_\ell \ve_{n\ell}^2 (G_\ell^\rR)^2 G_n^\rR G_n^\rA \left\{ v_\b^{(n)} A_\a^{n \ell} A_\g^{\ell n} + v_\g^{(n)} A_\a^{n \ell} A_\b^{\ell n} \right\} \nonumber \\
&\quad \quad \quad + \sum_\ell \ve_{n\ell}^2 (G_\ell^\rR)^3 G_n^\rA \left\{ v_\b^{(\ell)} A_\a^{n \ell} A_\g^{\ell n} + v_\g^{(\ell)} A_\a^{n \ell} A_\b^{\ell n} \right\} \nonumber \\
&\quad \quad \quad + \sum_ {m \ell} i\ve_{nm} \ve_{m\ell} \ve_{\ell n} A_\a^{nm} ( A_\b^{m\ell} A_\g^{\ell n} + A_\g^{m\ell} A_\b^{\ell n} ) (G_m^\rR)^2 G_\ell^\rR G_n^\rA 
\biggr]
\label{eq:band1} 
\\ 
&\equiv \sigma_{\a\b\g}^{\rm RRA, intra} + \sigma_{\a\b\g}^{\rm RRA, inter (1)} + \sigma_{\a\b\g}^{\rm RRA, inter (2)} + \sigma_{\a\b\g}^{\rm RRA, inter (3)} + \sigma_{\a\b\g}^{\rm RRA, inter (4)} 
\\
\sigma_{\a\b\g}^{\rm RA} &\equiv {\rm Im} \ {\rm tr} [ \hat{\cV}_\a (\hat{G}^\rR)^2 \hat{\cV}_{\b \g} \hat{G}^\rA ]
= 
{\rm Im} \sum_{nm} \sum_\k [ \lan |\hat{\cV}_\a | \ram (G_m^\rR)^2 \lam | \hat{\cV}_{\b\g} | \ran G_n^\rA ] 
\nonumber \\
&=
{\rm Im} \sum_{nm} \sum_\k (G_m^\rR)^2 G_n^\rA 
(\partial_\a \ve_n \delta_{nm} + i \ve_{mn} A_\a^{nm}) \nonumber \\
&\quad \times \left( \partial_\b \partial_\g \ve_n \delta_{nm} 
+ i [ \partial_\g \ve_{nm} A_\b^{mn} + (\b \leftrightarrow \g) ] 
+ i \ve_{nm} \partial_\g A_\b^{mn} 
- \sum_\ell [ \ve_{n\ell} A_\b^{m\ell} A_\g^{\ell n} + \ve_{m\ell} A_\g^{m\ell} A_\b^{\ell n} ] 
 \right) 
\nonumber \\
&=
{\rm Im} \sum_\k \sum_n \biggl[ (\partial_\a \ve_n) (\partial_\b \partial_\g \ve_n) 
- \sum_\ell \ve_{n\ell} (\partial_\a \ve_n) ( A_\b^{n\ell} A_\g^{\ell n} + A_\g^{n\ell} A_\b^{\ell n} ) ( G_n^\rR )^2 G_n^\rA \nonumber \\
&\quad \quad \quad + \sum_m \ve_{nm} [ \partial_\b \ve_{nm} A_\a^{nm} A_\g^{mn} + \partial_\g \ve_{nm} A_\a^{nm} A_\b^{mn}   ] ( G_m^\rR )^2 G_n^\rA \nonumber \\
&\quad \quad \quad + \sum_m \ve_{nm}^2 A_\a^{nm} \partial_\b A_\g^{mn} ( G_m^\rR )^2 G_n^\rA 
+ \sum_{m\ell} i \ve_{nm}  A_\a^{nm} [ \ve_{n\ell} A_\b^{m\ell} A_\g^{\ell n} + \ve_{m\ell} A_\g^{m\ell} A_\b^{\ell n} ] ( G_m^\rR )^2 G_n^\rA \biggr] .
\label{eq:band2}
\\
&\equiv \sigma_{\a\b\g}^{\rm RA, intra} + \sigma_{\a\b\g}^{\rm RA, inter (1)} + \sigma_{\a\b\g}^{\rm RA, inter (2)} + \sigma_{\a\b\g}^{\rm RA, inter (3)} + \sigma_{\a\b\g}^{\rm RA, inter (4)} 
\end{align}
In metals, terms $\propto \tau^2$ will be dominant in most cases. To check that the leading terms  are $\tau^2$, just use $G_n^\rR G_n^\rA = i\tau(G_n^\rR - G_n^\rA) = -2\tau {\rm Im} G^\rR$ which leads to $( G_n^\rR )^2 G_n^\rA \simeq 2i\tau^2 {\rm Im} G^\rR$, i.e. only imaginary parts are left. 

Let us first comment on the relation between the intraband terms $\sigma_{\a\b\g}^{\rm RRA,intra}$ and $\sigma_{\a\b\g}^{\rm RA,intra}$. Using integration by parts, we can show
\begin{align}
{\rm Im} \sum_{n\k}  v_\a^{(n)} v_{\b \g}^{(n)} (G_n^\rR)^2 G_n^\rA 
&= 
- {\rm Im} \sum_{n\k} v_\b^{(n)} v_{\g\a}^{(n)} (G_n^\rR)^2 G_n^\rA - 2 {\rm Im} \sum_{n\k}  v_{\a}^{(n)} v_\b^{(n)} v_\g^{(n)} (G_n^\rR)^3 G_n^\rA 
\nonumber \\
&=
{\rm Im} \sum_{n\k}  v_{\g}^{(n)} v_{\a\b}^{(n)} (G_n^\rR)^2 G_n^\rA 
=
{\rm Im} \sum_{n\k}  v_{\b}^{(n)} v_{\g\a}^{(n)} (G_n^\rR)^2 G_n^\rA .
\end{align}
Here, we put $v_\a^{(n)} \equiv \partial_\a \ve_n$ and $v_{\a \b}^{(n)} \equiv \partial_\a \partial_\b \ve_n$. This yields
\begin{align}
2\sigma_{\a\b\g}^{\rm RA, intra} = 2{\rm Im} \sum_{n\k}  v_\a^{(n)} v_{\b \g}^{(n)} (G_n^\rR)^2 G_n^\rA 
= 
- 2 {\rm Im} \sum_{n\k}  v_{\a}^{(n)} v_\b^{(n)} v_\g^{(n)} (G_n^\rR)^2 G_n^\rA = -\sigma_{\a\b\g}^{\rm RRA, intra}. 
\label{eq:intrarel}
\end{align}
Actually, these are the only intraband terms, namely $\sigma_{\a\b\g}^{\rm intra} =  ( \sigma_{\a\b\g}^{\rm RRA,intra} + \sigma_{\a\b\g}^{\rm RA,intra} )_{\tau^2} = \frac{1}{2} \sigma_{\a\b\g}^{\rm RRA,intra} =  - \sigma_{\a\b\g}^{\rm RA,intra} $. Note also that these terms reproduce those from the Boltzmann equation approach. 

Next, let us discuss the interband terms. Using $G_n^{\rR(\rA)} G_m^{\rR(\rA)} = (\ve_n - \ve_m)^{-1} (G_n^{\rR(\rA)} - G_m^{\rR(\rA)})$ and $G_n^{\rR} G_m^{\rA} = (\ve_n - \ve_m - 2i\gamma)^{-1} (G_n^{\rR} - G_m^{\rA})$, $\sigma_{\a\b\g}^{\rm RRA,inter (1)}$ and $\sigma_{\a\b\g}^{\rm RA,inter (1)}$ are calculated further as 
\begin{align}
\sigma_{\a\b\g}^{\rm RRA,inter (1)} + \sigma_{\a\b\g}^{\rm RA,inter (1)} 
= {\rm Im} \sum_\k \sum_{n\ell} v_\a^{(n)} ( A_\b^{n\ell} A_\g^{\ell n} + A_\g^{n\ell} A_\b^{\ell n} ) G_\ell^\rR G_n^\rA .
\end{align} 
Thus, we can verify that the interband  contributions $\propto \tau^2$, which are contained in $\sigma_{\a\b\g}^{\rm RRA,inter (1)}$ and $\sigma_{\a\b\g}^{\rm RA,inter (1)}$, cancel each other: $( \sigma_{\a\b\g}^{\rm RRA,inter (1)} + \sigma_{\a\b\g}^{\rm RA,inter (1)} )_{\tau^2} = 0$. The remaining term is the interband term in higher order of disorder. Similar calculations for $\sigma_{\a\b\g}^{\rm RRA,inter (2)}$, $\sigma_{\a\b\g}^{\rm RA,inter (2)}$, and $\sigma_{\a \b \g}^{\rm RRA,inter (3)}$ lead to 
\begin{align}
\sigma_{\a\b\g}^{\rm RRA,inter (2)} + \sigma_{\a\b\g}^{\rm RA,inter (2)} + \sigma_{\a \b \g}^{\rm RRA,inter (3)}
&= 2 \tau \sum_\k \sum_{n} [ v_\b^{(n)} \ve_{\a \g \delta} \Omega_\delta^{(n)} + v_\g^{(n)} \ve_{\a \b \delta} \Omega_\delta^{(n)} ] {\rm Im} G_n^\rR  \nonumber \\
&- {\rm Im} \sum_\k \sum_{n\ell} [ v_\b^{(n)} A_\a^{n \ell} A_\g^{\ell n} + v_\g^{(n)} A_\a^{n \ell} A_\b^{\ell n} ] G_\ell^\rR G_n^\rA  \nonumber \\
&- {\rm Im} \sum_\k \sum_{n\ell} [ v_\b^{(\ell)} A_\a^{n\ell} A_\g^{\ell n} + v_\g^{(\ell)} A_\a^{n\ell} A_\b^{\ell n} ] \ve_{n\ell} ( G_\ell^\rR )^2 G_n^\rA  \nonumber \\
&+ {\rm Im} \sum_\k \sum_{n\ell} [ A_\a^{n\ell} \d_\b A_\g^{\ell n} ] \ve_{\ell n}^2 ( G_\ell^\rR )^2 G_n^\rA\, .
\end{align}
The first term of the last equality is the so-called Berry curvature dipole term, which will be responsible for the nonlinear Hall effect (not considered here). Note that, for the two band system, two of the velocities $\ve_\ell$, $\ve_m$, and $\ve_n$ should be the same, leading to $\sigma_{\a\b\g}^{\rm RRA,inter (4)} = 0$. 

Let us derive Eq.~\eqref{eq:domQM} here. For the condition $\ve_{nm} \tau \ll 1$, we can expand the Green function up to the first order of $\ve_{\ell n}$ as $G_\ell^\rR \simeq G_n^\rR + \ve_{\ell n} (G_n^\rR )^2$. 
Noting that $|{\rm Re} [G_n^\rR] | \gamma < 1$, and 
\begin{align}
G_\ell^\rR G_n^\rA \simeq - 2 \tau {\rm Im} G_n^\rR + i\ve_{\ell n} \tau \{ (G_n^\rR)^2 + 2\tau {\rm Im} G_n^\rR \}, 
\end{align}
we can safely neglect $\sigma_{\a \b \g}^{\rm RRA,inter (3)}$ and $\sigma_{\a \b \g}^{\rm R(R)A,inter (4)}$ under the condition $\ve_{nm} \tau \ll 1$. If the Berry curvature dipole term is negligible, $\sigma_{\a \b \g}^{\rm inter}$ is calculated as 
\begin{align}
\sigma_{\a\b\g}^{\rm inter} &\simeq \sigma_{\a\b\g}^{\rm RRA,inter (1)} + \sigma_{\a\b\g}^{\rm RRA,inter (2)} + \sigma_{\a\b\g}^{\rm RA,inter (1)} + \sigma_{\a\b\g}^{\rm RA,inter (2)} + \sigma_{\a \b \g}^{\rm RA,inter (3)}
\nonumber \\ 
&= 2 \tau^2 \sum_\k \sum_{n\ell} {\rm Re} \left[ v_\a^{(n)} A_\b^{n\ell} A_\g^{\ell n} - v_\b^{(n)} A_\g^{n\ell} A_\a^{\ell n} - v_\g^{(n)} A_\a^{n\ell} A_\b^{\ell n} +  \frac{1}{2} \ve_{\ell n} A_\a^{n\ell} \partial_\b A_\g^{\ell n} + (\b \leftrightarrow \g) \right] \ve_{\ell n} {\rm Im} G_n^\rR .
\end{align}
Defining the symmetric tensor $g_{\a ;\b\g}^{n\ell} = {\rm Re} \left[ v_\b^{(n)} A_\g^{n\ell} A_\a^{\ell n} + v_\g^{(n)} A_\a^{n\ell} A_\b^{\ell n} - v_\a^{(n)} A_\b^{n\ell} A_\g^{\ell n} - \ve_{\ell n} A_\a^{n\ell} \partial_\b A_\g^{\ell n} /2  + (\b \leftrightarrow \g) \right]$, we arrive at Eq.~\eqref{eq:domQM}. 
\end{widetext}

\section{Relaxation time dependence at the vicinity of nodal points}
\label{sec:taudep}

\begin{figure}[h]
\includegraphics[width=85mm]{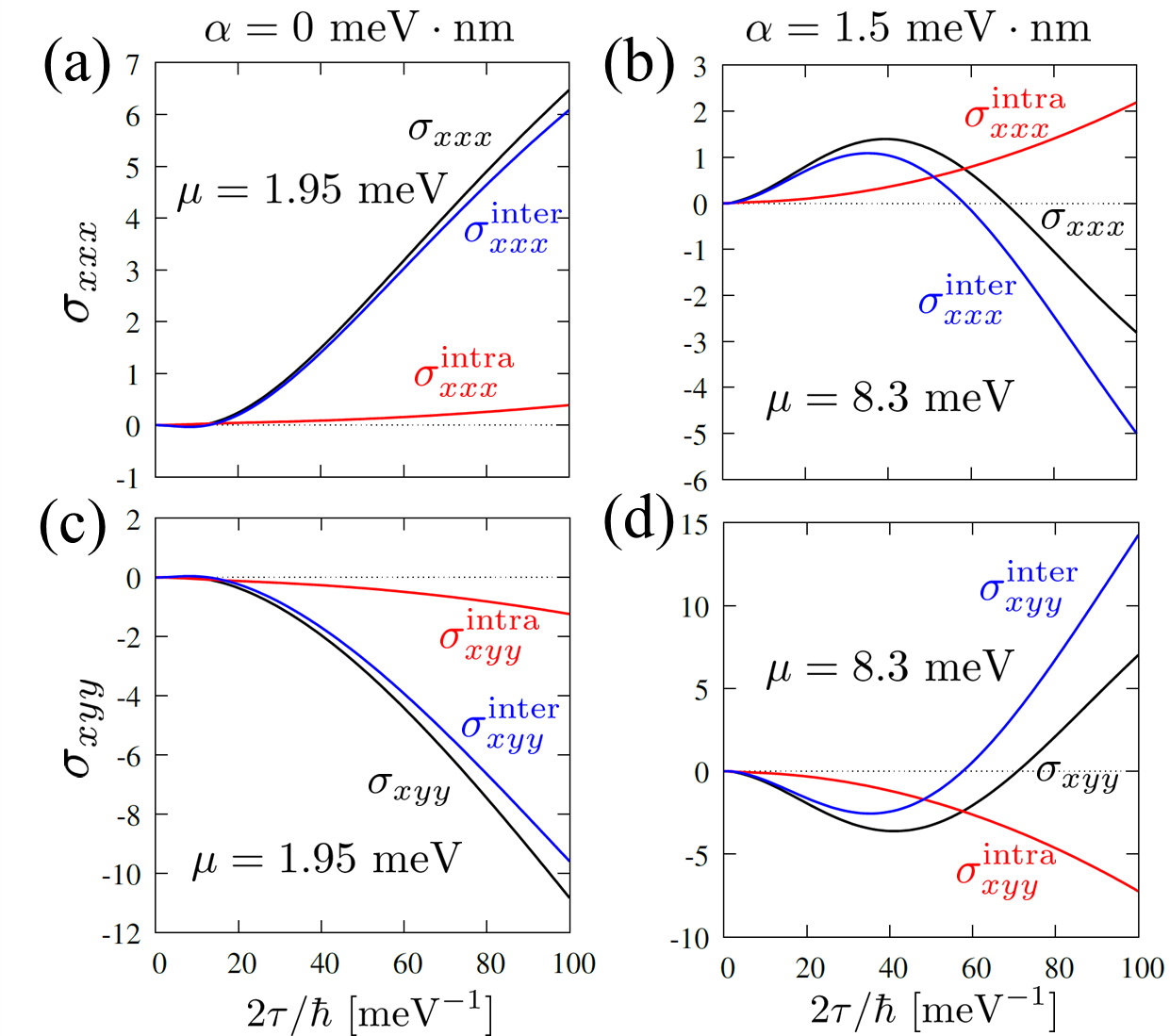}
\caption{ 
$\tau$ dependence of the second-order nonlinear conductivities [(a) and (b)] $\sigma_{xxx}$ and [(c) and (d)] $\sigma_{xyy}$ in cases of [(a) and (c)]  $\a = 0$ and [(b) and (d)]  $1.5$~$\rm meV \cdot nm$. 
Notations are same to Fig.~\ref{fig:2DGe100}(e) and Fig.~\ref{fig:2DGe110}(e). 
}
 \label{fig:taudep}
\end{figure}

Fig.~\ref{fig:taudep} shows the $\tau$ dependence of $\sigma_{xii}$ in 2DHG when the chemical potential $\mu$ is located near the nodal point. In such cases, the interband contribution is dominant even when $\tau$ is very large. This is because $\ve_{+-} \tau \ll 1$ is satisfied near the nodal point even when $\tau$ is large (see also Sec.~\ref{subsec:soi}). 

\end{document}